\documentstyle[prl,aps,epsf,rotate]{revtex}
\draft
\begin{document}
\twocolumn[\hsize\textwidth\columnwidth\hsize
     \csname @twocolumnfalse\endcsname

\title{Electron acceleration due to photon absorption:\\
A possible origin of the infinity problems in
relativistic quantum fields}

\author{W. A. Hofer} 
\address{Institut f\"ur Allgemeine Physik,
Technische Universit\"at Wien,
         A--1040 Vienna, Austria}

\maketitle


\begin{abstract}
Based on the concept of extended particles recently introduced
we perform a Gedankenexperiment accelerating single electrons
with photons of suitably low frequency. Accounting for relativistic
time dilation due to the acquired velocity and in infinite
repetition of single absorption processes it can be shown that
the kinetic energy in the infinite limit is equal to
$ m_{e} c^{2}/2 $. However, the inertial mass of the electron
seems enhanced, and it can be established
that this enhancement is described by the relativistic mass
effect. It appears, therefore, that although there exists a
singularity in interactions - the frequency required
to accelerate the particle near the limit of $c$ becomes
infinite - the energy of the particle itself approaches
a finite limit. Comparing with calculations of the
Lamb-shift by Bethe this result seems to provide the ultimate 
justification for the renormalization procedures employed.
\end{abstract} 


\vskip2pc]

A realistic interpretation of quantum theory (QT), or the interpretation
of the wave function as a real, i.e. {\em physical} wave, which
was Schr\"odinger's original concept \cite{HEI77},
has always been contradicted by compelling evidence.
Currently there exist four major obstacles for a realistic
reformulation of QT: (i) The uncertainty relations \cite{BRO88}, (ii)
the
dispersion relations for massive particles like electrons \cite{BRO25},
(iii)
intrinsic energy components due to electrostatic interactions, and
(iv) the experimental proofs against local and realistic theories
\cite{ASP82}.
The theoretical proofs against these theories by von Neumann or
Jauch and Piron \cite{NEU32,JAU63} are not as convincing,
since they leave quite a few loopholes, as Bohm and Bell pointed
out \cite{BOH66,BEL66}.

Due to this evidence the search for an interpretation of microphysics
in terms of {\em extended particles} remained something of a minority
program: although some efforts have been made, they always encountered
unsurmountable difficulties \cite{EXPAR}.

As recently established, these difficulties could be due to an only
restricted analysis of the fundamental relations in QT: this type of
analysis, called the {\em formal approach} proceeds from the
Schr\"odinger
relation and/or the commutation relations, which are accepted without
limits \cite{HOF98A}. Using a {\em realistic} approach, 
the result changes
drastically. If wave--features of single particles are interpreted
as physical waves describing the intrinsic structure due to particle
propagation, the fundamental axioms of QT gain the following meaning:
(i) Due to an additional and intrinsic energy component, the
dispersion relations of monochromatic waves are valid also for
massive particles, (ii) the Schr\"odinger equation is, also due to
this intrinsic energy component, no longer an exact equation,
(iii) the uncertainty relations remain valid, but they are no
longer axioms, but rather the error margin
due to the omission of intrinsic energy in
Schr\"odinger's equation. Especially the latter point
seemed interesting, because it allows, in principle, to describe
processes without a limit of precision also at the microphysical
level.

The last two problems to a realistic interpretation were
removed by establishing, that also the Maxwell equations
\cite{JAC84} are a
description of intrinsic particle properties, and while QT is
mainly concerned with longitudinal wave properties, electrodynamics
(ED) describes the transversal and intrinsic fields of propagation.
In this case it can be deduced that electrostatic fields of
interaction are a consequence of photon exchange, which removes
obstacle (iii), since these interactions vanish for a particle
in constant motion. In addition, the concept of spin in QT could
be referred to the transversal and intrinsic magnetic fields of
particle propagation: since in the realistic picture of extended
particles spin is an oscillating variable, a valid measurement
of spin correlations requires a local precision explicitly
higher than allowed by the uncertainty relations: in this case
the measurement cannot be described consistently within the
framework of QT, the experimental results (iv) therefore allow
no longer the conclusion, that a local and realistic theory
is inconsistent.
In view of these results it seems that the
approach provides an alternative to the standard interpretation,
and which can be used to describe fundamental processes on a
micro level. 

One of the most intriguing consequences of Einstein's theory of Special
Relativity (STR) \cite{EIN05A,EIN05B} is the energy relation for
a mass $m$ in a state of motion $u$ \cite{ADL87,OKU89}.

\begin{eqnarray}
E = \sqrt{m^2 c^4 + {\bf p}^2 c^2} = m \, \gamma \,c^2 \qquad
\gamma = \frac{1}{\sqrt{1 -  u^2 / c^2}}
\end{eqnarray}

In the small velocity range, commonly identified as the non--relativistic
range, the relation reduces to:

\begin{eqnarray}
E (u << c) = m c^2 + \frac{1}{2} \, m u^2
\end{eqnarray}

This expression is, as Einstein pointed out \cite{EIN06A}, equal to the 
classical expression in Newtonian mechanics with an additional term, the
rest energy of $m$. 

As recently established, the kinetic energy of an
electron is only half of its total energy, the correction results from
intrinsic energy components due to wave--like internal properties
\cite{HOF98A}. The result was obtained 
in a reference frame at rest, and it could also be derived that electrostatic
interactions correspond to photon emission and absorption processes. The
energy of a particle after absorption of a photon is given by:

\begin{eqnarray}
E_{1} = E_{0} + E_{ph} = m u_{0}^2 + m_{ph} c^2
\end{eqnarray}

where $m_{ph}$ denotes the mass of the absorbed photon. 
That photons or light pulses also transfer mass 
$m_{ph} = E _{ph} / c^2$ was demonstrated by Einstein in a 
Gedankenexperiment in 1906 \cite{EIN06A}. The connection of
photon mass with the rest energy term in STR is, also for 
a non--moving inertial frame, a rather obvious one:  
if the whole energy of a body is converted 
into electromagnetic energy, its mass shows up as mass of
the emitted photons.

Apart from this new interpretation, the result is no change of established
concepts and fairly standard. But as this result is derived within a 
non--relativistic framework, it is somewhat beyond Einstein's original
derivation, and it raises, for this reason, a new and interesting question.
In this case the problem becomes imminent, whether the relativistic energy
for moving electrons and its singularity at $u \rightarrow c$ do in
every case signify a real,
i.e. {\em physical} effect, or whether they cannot, in some cases, be considered
more or less {\em virtual}. Virtual meaning, that the effect only is a way
to account for energy transfer in electromagnetic fields. This question has
been addressed to some extent (see, for example, Adler's paper ref. \cite{ADL87}),
and the common understanding seems to be, that ''the time kept by a rapidly
moving particle is dilated and hence, as the particle's speed increases,
apparently greater time intervals are taken to produce the same effect, thus
the apparent increase in resistance.'' But in this case the same reasoning 
can be applied to the energy, and it must be analyzed, whether its
singularity is not, equally, a result of changed interaction characteristics.
To decide on this question we perform a Gedankenexperiment with
photons of (suitably) low frequency, and calculate the acceleration
of the electron due to photon absorption numerically.

The paper is organized as follows: first the one particle problem will
be treated in a frame at rest, and it is assumed, that the photon source is
in no other way interacting with the particle. The velocity and energy of
the particle due to photon absorption will be calculated and compared
with the relativistic expression. It will be shown, that the kinetic energy of a
particle cannot exceed the limit given by $ m c^2/2 $. As this
result suggests a thorougher analysis of the interaction process, we will
then assume that the classical and non--relativistic accelerations are due
to electrostatic fields. Comparing the classical changes of
velocity to the changes pertaining to the photon--absorption
model, it can be established that the relativistic mass effect
- or the enhancement of inertia - must be attributed
to the time dilation in the electron system.
The result is compared to
existing measurements of the ratio $(e / m) = (e / m) (u) $ 
\cite{KAU01,KAU02,KAU03,KAU05,KAU06,NEU14,GUY15,CUS81},
and  it is shown that the interpretation is in
accordance with experimental evidence.   
Relating the derived results on
electron energy to the prevailing infinity problems in quantum
electrodynamics and their proposed solution via
renormalization procedures \cite{FER32}--\cite{WEI49} it can be
shown that these infinity problems typically arise from the 
singular properties of energy in relativity theory.

The electron with mass $m$ shall initially be at rest. The photon
source with frequency $\omega_{0}$ in our inertial frame, where $\omega_{0}$
shall be very small in order to yield differential--like accelerations, is
also at rest in the same frame $S_{0}$. We consider only linear accelerations,
the system is therefore limited to one dimension, say  $x$, with the initial
position of the particle $x_{0} = 0$. As the acceleration during interaction
depends on the differential of energy density, as recently demonstrated \cite{HOF98A},
we shall not consider the acceleration process but merely evaluate the velocity
before and after absorption.

The only effect to be considered in the one--particle problem is the
relativistic time dilation due to relative motion of the particle.
The energy in absorption processes depends on the eigenzeit
in the system of the particle \cite{HOF98A}. Due to time
dilation the absorbed energy will be diminished, energy
and velocity of the particle after $n$ photon absorptions
are therefore given by:

\begin{eqnarray}\label{004}
E_{n} & = & \hbar \omega_{0} \left[1 + \sum\limits_{i=0}^{n-1}
\sqrt{1 - \left( \frac{u_{i}}{c} \right)^{2}} \right] = m \, u_{n}^2
\nonumber \\
u_{n} & = & \sqrt{\frac{\hbar \omega_{0}}{m}} \left[1 + \sum\limits_{i=0}^{n-1}
\sqrt{1 - \left( \frac{u_{i}}{c} \right)^{2}} \right]^{1/2}
\end{eqnarray}

\begin{figure}
\epsfxsize=0.9\hsize
\epsfbox{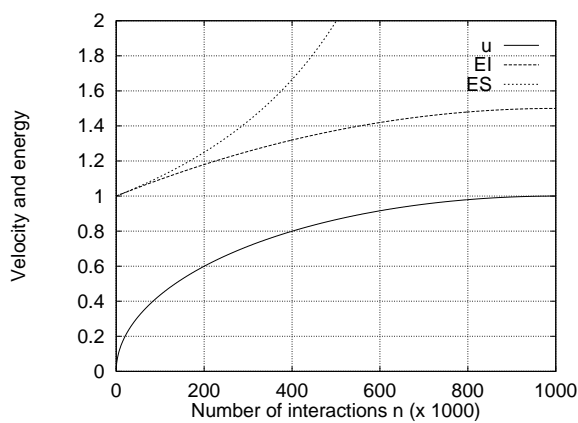}
\vspace{0.2 cm}
\caption{
Energies in the interaction ($EI$) and STR ($ES$) model. All units in
the numerical calculation were relativistic units
($m = 1, c = 1, [u] = c, \hbar \omega_{0} = 4 \times 10^{-6}$).
The kinetic energy in the interaction model converges to a final value 
of $m c^2/2 $, while in the STR model it becomes irregular for
$u \rightarrow 1$.}
\label{fig001}
\end{figure}

To compare with Einstein's expression we add the rest energy
and take only half of the total energy of the particle. Performing the interactions
with a suitably chosen photon frequency in a numerical calculation, it can be
seen that the kinetic energy of the particle is equal to Einstein's expression only in the
low velocity range, while in the high range the energy converges to a limit of
$1.5\, m c^2$ in the interaction model, whereas it becomes irregular in the STR model
(Fig. \ref{fig001}).

This result for the energy of the particle suggests a comparison with
accelerations in electrostatic fields, since the energy expression in
relativistic physics can only be accounted for, if the 
{\em resistance} of a particle to change its state of motion
is due to the difference between the classical concept and the concept
of photon interactions.

Let an electrostatic potential $U_{0}$ exist in the system, which is
the potential difference between $x_{0} = 0$ and $x_{1} = L$. The classical model
assumes the transfer of energy from the electrostatic field to the
particle to be independent of the particle's state of motion. This is,
in the interaction model, equal to no frequency changes of the photons
absorbed, so the classical velocity difference $ \triangle u_{n}^c $ between 
two interaction processes will be:

\begin{eqnarray}
E_{n} &=& \frac{m}{2} u_{n}^2 = \frac{n}{2} \,\hbar \omega_{0} 
\nonumber \\
\triangle u_{n}^c &=& \sqrt{\frac{\hbar \omega_{0}}{m}} \left(
\sqrt{n} - \sqrt{n-1} \right)
\end{eqnarray}

The velocity difference  $\triangle u_{n}^r $  due to time
dilation in the electron system is significantly different:

\begin{eqnarray}
\triangle u_{n}^r = 
u_{n} - u_{n-1}
\end{eqnarray}

where the velocities $ u_{n} $ are given by Eq. \ref{004}. The accelerations
in the electrostatic field therefore cannot be independent of the particle's
state of motion. If the deviation from the classical behavior, inevitably
bound to occur, is interpreted as an increase of inertial mass, it must be
described by a variable $ \alpha' (u) $, defined by the following
relations ($N \triangle x = L$):

\begin{eqnarray}
\triangle u_{n}^c = \frac{1}{m}\frac{U_{0}}{N} \frac{\triangle t}
{\triangle x} \quad
\triangle u_{n}^r = \frac{1}{m \alpha'}\frac{U_{0}}{N} \frac{\triangle t}
{\triangle x}    
\end{eqnarray}

In addition it has to be considered that the classical theory
miscalculates the already acquired velocity of the electron.
The variable $ \alpha' $ therefore has to be corrected with
the ratio of velocities $ u_{n}^c/u_{n}^r $.
Then the variable $ \alpha $, describing the virtual change of
inertia due to time dilation, can be calculated, it will be:

\begin{eqnarray}
\alpha = \alpha' \,\frac{u_{n}^c}{u_{n}^r} =
\sqrt{\frac{\hbar \omega_{0}}{m}} \,
\frac{\sqrt{n} - \sqrt{n-1}}{u_{n} - u_{n-1}} 
\frac{u_{n}^c}{u_{n}^r}
\end{eqnarray}

\begin{figure}
\epsfxsize=0.9\hsize
\epsfbox{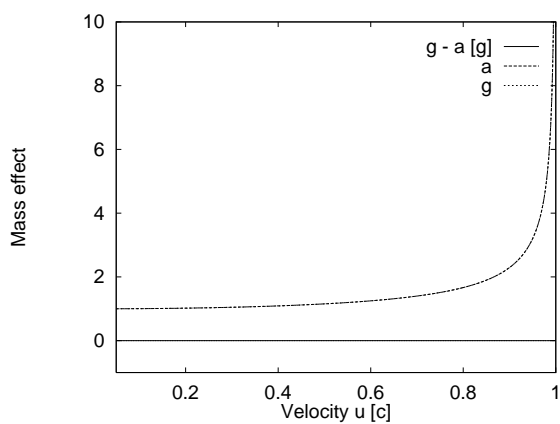}
\vspace{0.2 cm}
\caption{Mass effect due to frequency changes of photons. The
frequency shifts due to velocity of the particle lead to an
observed but virtual increase of inertial mass. The difference
between $\gamma$ and $\alpha$ (g and a) is insignificant 
over the whole velocity range from 
$u = 0.05$ to $u \approx 0.99 c$.}
\label{fig002}
\end{figure}

The numerical calculation has been performed with an identical dataset 
and the result of this calculation is displayed in Fig.
\ref{fig002}. It can be established that the effect in this
case is observable, but not a real, i.e. {\em physical} effect.
The mass of the particle remains constant over the whole range of
velocities. But due to the classical conception of electrostatic
interactions decreasing accelerations in the high velocity range
must be interpreted in some other way: in special relativity they
are interpreted as an increase of inertial mass. As this usage of
the term mass is a little anachronistic, we wish to emphasize that
it is to be understood in the sense of Bondi \cite{BON79}, who 
related inertia, or resistance to the total mass of a particle:  
''After all, the mass of a moving body can be taken to be either
its rest mass or its total mass which includes the mass of its
kinetic energy.''
While the relativistic mass formulas in Einstein's theory are
''artifacts of the kinematical transformation of space and time'' 
\cite{ADL87}, and a result of real changes in the structure of
the electron in Lorentz' theory \cite{LOR04}, in the present context
they are due to the changed characteristics of interaction described
by the time dilation in moving reference frames.

The most extensive measurements of the ratio
$(e / m)$ of electrons were performed in the first fifteen years of this
century by Kaufmann, Neumann, and Guye and Lavanchy \cite{KAU06,NEU14,GUY15}.
The motivation for this extensive research was the problem of electromagnetic
mass of the electron described by Abraham's theory \cite{ABR03}. The 
experiments were performed to decide between Abraham's and Einstein's or
Lorentz' theory of electrons \cite{ABR03,EIN05A,LOR04}, and while initial 
results seemed to favor Abraham, the question was
finally settled by Neumann in favor of Einstein or
Lorentz \cite{NEU14}. Neumann's result was confirmed by Guye
and Lavanchy \cite{GUY15}. As the result obtained in the interaction
model of electrostatic interaction is equal, but for numerical artifacts
(in the order of 10$^{-4}$ from 0.05 to 0.99 c, and which decrease with 
decreasing photon frequencies and thus smoother
accelerations) to Einstein's or Lorentz results, it is fully compatible
with the experimental evidence. But as this result leaves the electron
energy finite even in the limit $ u \rightarrow c $, it raises 
questions in relativistic quantum field theories which are related
to the notorious infinity problems in this area.

Following Weisskopf in his treatment of the free electron 
\cite{WEI39}, there are two infinite contributions to the 
self energy of an electron: 
(i) The electrostatic energy diverging with the radius $a$ of
the electron, and (ii) the energy due to vacuum fluctuations
of the electromagnetic fields. The two energies $W_{st}$ and
$W_{fluct}$ have been calculated by Weisskopf as:

\begin{equation}
W_{st} = \lim_{a \rightarrow 0} \, \frac{e^2}{a} \qquad
W_{fluct} = \lim_{a \rightarrow 0} \, 
\frac{e^2 h}{\pi m c a^2}
\end{equation}

As shown previously, the electrostatic contribution vanishes
in the context of the present theory, because electrostatic
interactions can be referred to an exchange of photons 
\cite{HOF98A}: for an electron in constant motion and, 
especially, an electron at rest, the electrostatic fields
of interaction vanish. As does the electrostatic contribution
to the infinite self-energy of the electron.

In the first calculation to master the infinity problems of 
quantum electrodynamics Bethe derived the following expression
for the Lamb shift of the hydrogen electron in an 
s-state \cite{BETH47}:

\begin{equation}
W_{ns}' = C \cdot \ln \frac{K}{\langle E_{n} - E_{m} \rangle_{AV}}
\end{equation}

where C is a constant, 
$ \langle E_{n} - E_{m} \rangle_{AV}$ the average
energy difference between states $m$ and $n$,
and K determined by the cutoff of 
electromagnetic field energy. The prime refers to mass
renormalization, since the - infinite - contribution to
the electron energy due to electrostatic mass is 
subtracted. The second infinity, the infinity of vacuum
fluctuations, is discarded by defining the cutoff K, which
in Bethe's calculation is equal to $m c^2$. 
But while the energy of
the field could have any value, if the actual energy of
the electron has a singularity at $u = c$
(and K could therefore be infinite), this is not the
case if the energy remains finite in this limit: in this case
the total energy difference between a relativistic electron and
an electron at rest is $m c^2$ according
to our calculations. This is, incidentally,
equal to the ''rest energy'' of the electron. It seems,
therefore, that the renormalization procedures, 
a common practice now for quite some time 
\cite{SCH94}, may have their 
ultimate justification in finite electron energy as well as
vanishing electrostatic energy components.

I'd like to thank Tom van Flandern, whose
remarks on possible consequences of finite propagation
velocities made me rethink the question of relativistic
energies from this angle.


%
%

\begin{references}
\bibitem{HEI77}
W. Heisenberg ''Remarks on the Origin of Uncertainty'' in
W. C. Price (ed.) {\it The Uncertainty Principle and the
Foundations of Quantum Mechanics}, Wiley \& Sons, New York (1977)
\bibitem{BRO88}
L. de Broglie {\it Les Incertitudes d'Heisenberg et l'Interpre-} \\
{\it tation Probabiliste de la Mecanique Ondulatoire},
Gauthier Villars, Paris (1988)
\bibitem{BRO25}
L. de Broglie {\it Ann. Phys.} \, {\bf 3}, 22 (1925)
\bibitem{ASP82}
A. Aspect, J. Dalibard, and G. Roger \,{\it Phys. Rev. Lett.} \,
{\bf 49}, 1804 (1982)
\bibitem{NEU32}
J. von Neumann {\it Mathematische Grundlagen der Quantenmechanik},
Springer, Berlin (1932)
\bibitem{JAU63}
J. M. Jauch and C. Piron, {\it Helv. Phys. Acta} \, {\bf 36}, 827 (1963)
\bibitem{BOH66}
D. Bohm and J. Bub {\it Rev. Mod. Phys.} \,{\bf 38}, 470 (1966)
\bibitem{BEL66}
J. S. Bell, {\it Rev. Mod. Phys.} \,{\bf 38}, 447 (1966)
\bibitem{EXPAR}
the most important results are contained in the following
papers:
H. Poincare {\it Rend. Circ. Mat. Palermo} \, {\bf 21}, 129 (1906);
J. Frenkel {\it Z. Physik} \, {\bf 32}, 518 (1925);
E. C. G. Stueckelberg {\it Comptes Rendu} \, {\bf 207}, 387 (1938);
F. Bopp {\it Ann. Physik} \, {\bf 38}, 345 (1940);
A. Pais {\it Verh. Kon. Ned. Akad. Nat.} \, {\bf 19}, 1 (1946);
S. Sakata {\it Prog. Theor. Phys.}\, {\bf 2}, 145 (1947);
\bibitem{HOF98A}
W. A. Hofer, {\it Physica A} {\bf 256}, 178 (1998)
\bibitem{JAC84}
J. Jackson {\em Classical Electrodynamics}, Wiley \& Sons,
New York (1984)
\bibitem{EIN05A}
A. Einstein  {\it Ann. Physik}\, {\bf 17}, 891 (1905)
\bibitem{EIN05B}
A. Einstein  {\it Ann. Physik} \, {\bf 18}, 639 (1905)
\bibitem{ADL87}
C. G. Adler,  {\it Am. J. Phys.} \, {\bf 55}, 739 (1987)
\bibitem{OKU89}
L. B. Okun,  {\it Phys. Today} \,{\bf 43}(6), 31 (1989)
\bibitem{EIN06A}
A. Einstein  {\it Ann. Physik} \,{\bf 20}, 371 (1906)
\bibitem{KAU01}
W. Kaufmann,  {\it Nachr. K. Ges. Wiss. Goettingen} \,{\bf 2}, 143 (1901)
\bibitem{KAU02}
W. Kaufmann  {\it  Nachr. K. Ges. Wiss. Goettingen} \,{\bf 3}, 291 (1902);
{\it Phys. Zeitschr.}\, {\bf 4}, 54 (1902)
\bibitem{KAU03}
W. Kaufmann  {\it  Nachr. K. Ges. Wiss. Goettingen} \, {\bf 4}, 90 (1903)
\bibitem{KAU05}
W. Kaufmann,  {\it Sitzungsber. K. Preuss. Akad. Wiss.} \, {\bf 2}, 949 (1905)
\bibitem{KAU06}
W. Kaufmann  {\it Ann. Physik} \, {\bf 19}, 487 (1906)
\bibitem{NEU14}
G. Neumann  {\it Ann. Physik} \, {\bf 28}, 529 (1914)
\bibitem{GUY15}
C. E. Guye  and C. Lavanchy  {\it Compt. Rend.} \, {\bf 161}, 52 (1915)
\bibitem{CUS81}
for a complete survey as well as the theoretical background of these measurements
see J. T. Cushing,  {\it Am. J. Phys.}\, {\bf 49}, 1133 (1981)
\bibitem{FER32}
E. Fermi  {\it Rev. Mod. Phys.}\,{\bf 4}, 87 (1932)
\bibitem{BLO37}
F. Bloch  and A. Nordsieck  {\it Phys. Rev.} \, {\bf 52}, 54 (1937)
\bibitem{KRA58}
H. Kramers  {\it Collected Scientific Papers}, Amsterdam (1958)
\bibitem{PAU39}
M. Fierz  and W. Pauli  {\it Proc. Roy. Soc. London A}\, {\bf 173},
211 (1939)
\bibitem{WEI39}
V. F. Weisskopf  {\it Phys. Rev.}\, {\bf 56}, 72 (1939)
\bibitem{WEN43}
G. Wentzel  {\it Einf\"uhrung in die Quantentheorie der
Wellenfelder}, Vienna (1943)
\bibitem{BETH47}
H. A. Bethe  {\it Phys. Rev. } \, {\bf 72}, 339 (1947)
\bibitem{SCHW48}
J. Schwinger  and  V. F. Weisskopf  {\it Phys. Rev.}\, {\bf 73},
1272 A (1948)
\bibitem{SCHW58}
J. Schwinger  {\it Selected Papers on Quantum Electrodynamics},
New York (1958)
\bibitem{FEY48}
R. P. Feynman  {\it Phys. Rev.} {\bf 74}, 939 ; 1430 (1948)
\bibitem{WEI49}
V. F. Weisskopf  {\it Rev. Mod. Phys.}\, {\bf 21}, 305 (1949)
\bibitem{BON79}
{\it Einstein: A Centenary Volume}, A. P. French  (ed.) Cambridge MA (1979)
p. 127
\bibitem{LOR04}
H. A. Lorentz  {\it Proc. R. Acad. Sci. Amsterdam} \, {\bf 6}, 809 (1904)
\bibitem{ABR03}
M. Abraham  {\it Ann. Physik} \, {\bf 10}, 105 (1903)
\bibitem{SCH94}
for a fairly recent account of renormalization see 
S. Schweber, {\it QED and the Men Who Made It}, Princeton
(1994)
\end{references}
\end{document}